\documentclass[aps, prb, 10pt, twocolumn, superscriptaddress, longbibliography]{revtex4-1}
\usepackage{amsmath,amsfonts,amssymb}
\usepackage{hyperref}
\usepackage{graphicx}
\usepackage{bm}
\usepackage{esint}
\usepackage{xcolor}

\begin{document}

\title{Linked skyrmions in shifted magnetic bilayer}

\author{Sumit Ghosh}
\email{ghoshs@fzu.cz}
\affiliation{The Institute of Physics of the Czech Academy of Sciences, 162 00 Prague, Czech Republic}
\affiliation{Peter Gr{\"u}nberg Institut (PGI-1), Forschungszentrum J{\"u}lich GmbH, 52425 J{\"u}lich, Germany}

\author{Hiroshi Katsumoto}
\affiliation{Peter Gr{\"u}nberg Institut (PGI-1), Forschungszentrum J{\"u}lich GmbH, 52425 J{\"u}lich, Germany}

\author{Gustav Bihlmayer}
\affiliation{Peter Gr{\"u}nberg Institut (PGI-1), Forschungszentrum J{\"u}lich GmbH, 52425 J{\"u}lich, Germany}

\author{Moritz Sallermann}
\affiliation{Peter Gr{\"u}nberg Institut (PGI-1), Forschungszentrum J{\"u}lich GmbH, 52425 J{\"u}lich, Germany}
\affiliation{Institute for Theoretical Physics, RWTH Aachen University, 52056 Aachen, Germany}
\affiliation{Science Institute and Faculty of Physical Sciences, University of Iceland, 107 Reykjav\'{i}k, Iceland} 

\author{Vladyslav M. Kuchkin}
\affiliation{Department of Physics and Materials Science, University of Luxembourg, L-1511 Luxembourg, Luxembourg}

\author{Filipp N. Rybakov}
\affiliation{Department of Physics and Astronomy, Uppsala University, SE-75120 Uppsala, Sweden}

\author{Olle~Eriksson}
\affiliation{Department of Physics and Astronomy, Uppsala University, SE-75120 Uppsala, Sweden}

\author{Stefan Bl\"ugel}
\affiliation{Peter Gr{\"u}nberg Institut (PGI-1), Forschungszentrum J{\"u}lich GmbH, 52425 J{\"u}lich, Germany}
\affiliation{Institute for Theoretical Physics, RWTH Aachen University, 52056 Aachen, Germany}

\author{Nikolai S. Kiselev}
\affiliation{Peter Gr{\"u}nberg Institut (PGI-1), Forschungszentrum J{\"u}lich GmbH, 52425 J{\"u}lich, Germany}

\begin{abstract}
Magnetic solitons have recently attracted significant attention due to their intricate physical properties and potential applications in information processing. The majority of the studies in this field, however, are focused on a particular type of solitons known as skyrmions, characterised by a unit topological charge. Here, we present a shifted magnetic bilayer that can demonstrate magnetic solitons with arbitrary large topological charges. These configurations, which we call \textit{linked skyrmions}, consist of multiple skyrmions linked together with topological point defects. These topological point defects, termed as \textit{anti-aligned} points, originate from the mutually orthogonal Dzyaloshinskii-Moriya interaction in two different layers. By tuning the interlayer exchange coupling and the external magnetic field, one can also achieve different ground states in this bilayer. Additionally, the system also demonstrates conventional \textit{skyrmion-bags} and $k\pi$-skyrmions. Finally we propose a suitable material candidate where these magnetic configurations can be realised. Our findings, thus, can provide a way to achieve solitons with large topological charge and realise them in realistic systems.
\end{abstract}

\maketitle

\section{Introduction}

Magnetic solitons \cite{Kosevich1990, Manton2004} are localised magnetic textures which behave like classical particles and can be characterised by a topological index. Their complex mathematical and physical features have stimulated a large number of studies in recent years leading to the discovery of unconventional magnetic texture both in two dimensions \cite{Lin2015, Foster2019, Rybakov2019, Gobel2021} and three dimensions \cite{Okumura2020, Kent2021, Azhar2022, Zheng2023, Liu2024}. They can be moved with external charge current \cite{Woo2016, Wang2019} which makes them a fitting candidate for data storage devices \cite{Fert2013, Yu2016, He2023}. In this regard, the two-dimensional configurations are more promising \cite{ Fert2017, Everschor-Sitte2018} due to their compatibility with current device architecture.

In condensed matter system, the most common example of magnetic solitons are the \textit{skyrmions} \cite{Rossler2006, Nagaosa2013} which is characterised by an integer topological index.
It can be calculated as below for continuous systems,
\begin{eqnarray}
Q = \frac{1}{4\pi} \int  dx \ dy \  \mathbf{m}\cdot[\partial_{x}\mathbf{m}\times\partial_{y}\mathbf{m}],
\label{Q}
\end{eqnarray}
where $\mathbf{m}$ is the unit vector denoting the magnetisation direction, whereas for discrete systems, the simplicial method applies~\cite{Berg1981, Mapping_degree_theory}. This is also known as the topological charge of the corresponding magnetic texture or simply the skyrmion charge. While most of the studies are focused on the skyrmions with $|Q|$=1, recently composite skyrmions with arbitrary large topological charge \cite{Rybakov2019, Foster2019}, commonly known as skyrmion bags, have also been observed experimentally \cite{Tang2021, Powalla2023, YZhang2024, Yang2024, Hassan2024}. These complex magnetic textures arise from the competition of different magnetic interactions among which the most prominent ones are the Heisenberg exchange and the Dzyaloshinskii-Moriya interaction (DMI) \cite{Dzialoshinskii1957, Moriya1960}. The interplay between these interactions along with an external magnetic field can result in different types of magnetic solitons.

\begin{figure}[t!]
\centering
\includegraphics[width=0.45\textwidth]{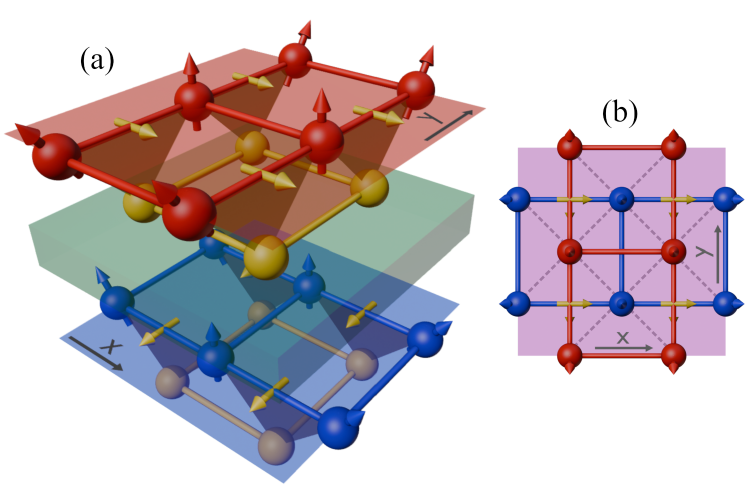}
\caption{Schematic of the shifted magnetic bilayer. (a) The red, blue and yellow spheres represent the top magnetic, bottom magnetic and non-magnetic sites. The resulting directions of the Dzyaloshinskii-Moriya vectors are shown with yellow arrows. Red and blue arrows show the favourable orientation of the spin due to the Dzyaloshinskii-Moriya interaction. The green region shows the nonmagnetic spacer layer. (b) Top view of the bilayer. The dashed lines show the nearest neighbours from opposite layers.
}
\label{fig:bilayer}
\end{figure}

In this paper, we propose a class of complex magnetic solitons with arbitrarily large topological charge which can be realised in a shifted magnetic bilayer. Our model consists of two magnetic layers where the DMI in each layer is mutually perpendicular to each other (Fig.~\ref{fig:bilayer}). Such configuration can be achieved by the proximity of a nonmagnetic layer with suitable spin-orbit coupling (SOC). The interlayer coupling can be controlled by using a suitable nonmagnetic spacer. With such anisotropic DMI and an external magnetic field, the proposed bilayer can demonstrate a wide variety of complex solitons along with the existing bag-like skyrmions. By analysing its phase space, we show the suitable parameter region to obtain complex textures with higher topological charges and establish their topological stability with homotopy group analysis. Finally, through a systematic first-principles study, we propose a suitable combination of materials where such complex solitons can be observed.   


\begin{figure*}
\centering
\includegraphics[width=\linewidth]{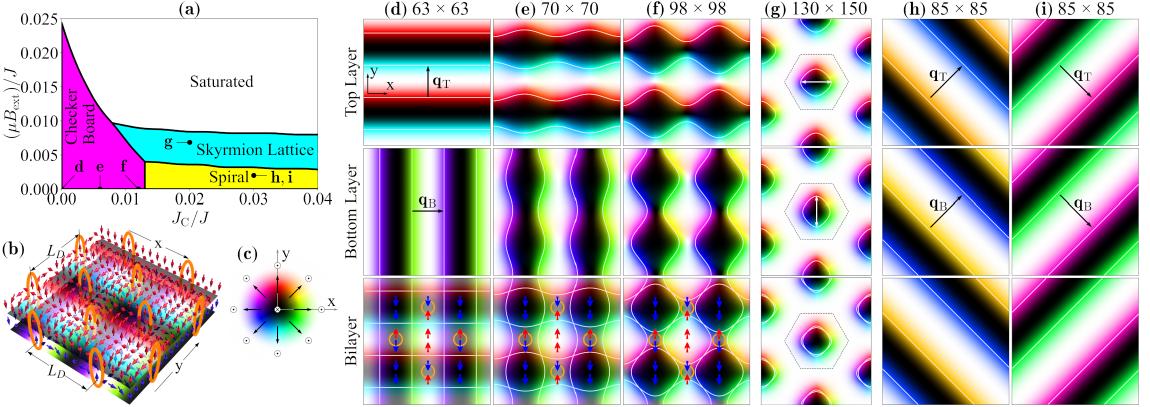}
\caption{Phase diagram and different periodic ground states of the bilayer Hamiltonian (Eq.~\ref{H}). (a) Phase diagram showing checkerboard (magenta), spiral (yellow), skyrmion lattice (cyan), saturated ferromagnetic (white) phase, and the location of the representative configurations. (b) Schematic of the checkerboard in the decoupled limit ($J_C$=0) and in the absence of the external magnetic field ($B_\mathrm{ext}$=0). Orange ellipses show the \textit{anti-aligned} points. (c) Colour code used in the figures demonstrated with an isotropic N{\'e}el skyrmion. $\odot(\otimes)$ shows the region with positive(negative) out of plane components denoted by black(white) colour. Layer resolved checkerboard pattern with $B_\mathrm{ext}$=0.0 and (d)$J_C$=0.0$J$, (e)$J_C$=0.006 and (f)$J_C$=0.012$J$. $\bm{q}_\mathrm{T,B}$ shows the direction of the spiral $q$ vector in top,bottom layer. The numbers on the top show the optimal cell size. White lines show the trajectory of the zero out-of-plane magnetisation. Orange circles show the location of the \textit{anti-aligned} points, and red and blue arrows show the out-of-plane magnetic alignment from the top and bottom layer in the corresponding region. (g) Skyrmion lattice phase obtained at ($J_C$=0.02$J$, $\mu B_\mathrm{ext}$=0.007$J$). The hexagonal region denotes the domain of the skyrmion (Eq.~\ref{Q}) and the white arrows show the elongation of the skyrmion in individual layers. (h,i) Show two degenerate spin spiral configurations obtained at ($J_C$=0.03$J$, $\mu B_\mathrm{ext}$=0.002$J$).
}
\label{fig:phase}
\end{figure*}

\section{Results and Discussion}


Our system is composed of two magnetic layers made of square lattices (the lattice constant is chosen to be 1) such that the centre of the top and bottom layers are shifted by (1/2,1/2) (Fig.~\ref{fig:bilayer}) like a Zincblende structure. A square lattice of nonmagnetic material as a source of SOC is placed such that the centre of this lattice is shifted by (1/2,0) from the bottom layer and (0,1/2) from the top layer (Fig.~\ref{fig:bilayer}). The structural symmetry gives rise to DMI along specific directions only \cite{Fert1980, Fert2013} as shown in Fig.~\ref{fig:bilayer}. There might be a small induced DMI in the other direction and along the interlayer bonds which, without loss of generality is set to zero. For simplicity, we also ignore the effect of any magneto-crystalline anisotropy which does not change the qualitative nature of the outcomes. The direction of the dominant DMI vector ($\bm{d}_{ij}$) along the line $\bm{r}_{ij}$ connecting site $i$ and $j$  for top($\mathrm{T}$) and bottom($\mathrm{B}$) layers are defined as

\begin{align}
\bm{d}^\mathrm{T}_{ij}= 
    \begin{cases}
        0, \!\!&\bm{r}_{ij}\! \parallel \!\bm{e}_x,\\
        \bm{e}_x, \!\!&\bm{r}_{ij}\! \parallel \!\bm{e}_y,
    \end{cases}
\,\,\,\,\bm{d}^\mathrm{B}_{ij}= 
    \begin{cases}
        -\bm{e}_y, \!\!&\bm{r}_{ij}\! \parallel \!\bm{e}_x,\\
        0, \!\!&\bm{r}_{ij}\! \parallel \!\bm{e}_y,
    \end{cases}
\label{DMI-vector}
\end{align}
and the complete lattice magnetic Hamiltonian for the bilayer is given by
\begin{eqnarray}
&\mathcal{H} =& -J\sum_{\langle ij\rangle;l} \bm{m}^{l}_i \cdot \bm{m}^{l}_j - D\!\sum_{\langle ij\rangle;l} \bm{d}^l_{ij} \!\cdot\! ( \bm{m}^{l}_i \times \bm{m}^{l}_j )   \nonumber \\
& & -J_\mathrm{C} \sum_{\langle i'j' \rangle}\bm{m}^\mathrm{T}_{i'} \cdot \bm{m}^\mathrm{B}_{j'} - \mu B_\mathrm{ext} \sum_{i;l} \bm{e}_z \cdot \bm{m}^l_i,
\label{H}
\end{eqnarray}
where $\langle ij\rangle$ indicates summation over the nearest neighbours in the same layer and $\langle i'j' \rangle$ corresponds to the summation over the nearest neighbours of different layers. The superscript $l$ runs over the top($\mathrm{T}$) and bottom ($\mathrm{B}$) layer. $J$ and $J_C$ correspond to the Heisenberg exchange parameters within the same and between different layers, respectively, and $D$ is the magnitude of DMI. Here we consider $J,J_C>$0 resulting in a ferromagnetic coupling and $D$=0.2$J$. We consider the same $J,J_C$ along both $x$ and $y$ direction, which is a valid assumption for $J_C<<J$. The model, therefore, can be considered as two mono-axial chiral magnets \cite{Kishine2015} coupled by the ferromagnetic interlayer exchange ($J_C$). $\mu$ is the onsite magnetic moment and $B_{ext}$ is the external magnetic field applied along $z$ axis.

\begin{figure*}[ht!]
\centering
\includegraphics[width=\linewidth]{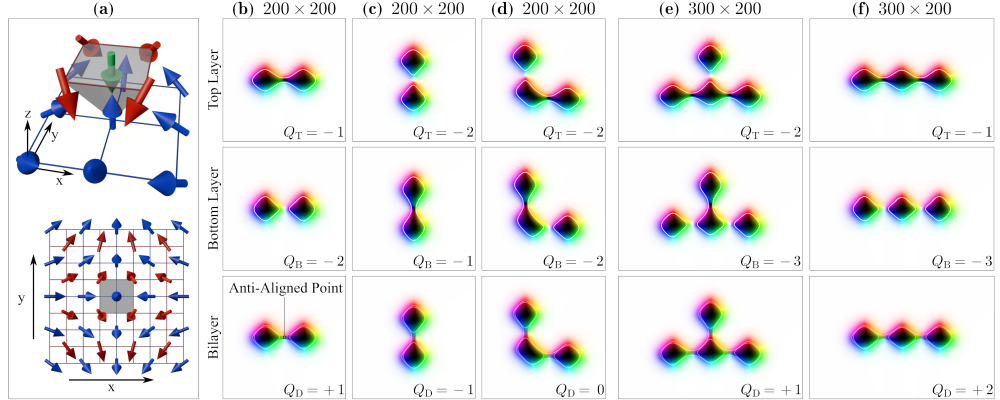}
\caption{
Linked skyrmion with different topological charges. (a) Enlarged view of the topological point defect. The red and blue arrows show the magnetic moment on the top and bottom layers. The green arrow shows the average of the four magnetic moments of the top layer, which constitutes the \textit{anti-aligned} point confined within the grey region. The bottom panel shows the top view of the extended region containing the point defect, where the grey box denotes the region of the \textit{anti-aligned} point. (b-f) Different linked skyrmions with their topological charge from the top ($Q^\mathrm{T}$) and bottom ($Q^\mathrm{B}$) layers, and the corresponding charge of the topological point defect ($Q^\mathrm{D}$). All configurations are stabilised at ($J_C$=0.02$J$, $\mu B_\mathrm{ext}$=0.007$J$) in a 400$\times$400 mesh. The numbers on top denote the dimensions of the plotting regions. The small square region of the bottom panel of column (\textbf{b}) corresponds to the configuration at the bottom panel of column (\textbf{a}).  
}
\label{fig:link}
\end{figure*}

\subsection{Phase diagram}

The phase diagram of the system is constructed by varying the interlayer coupling $J_C$ and external field $B_\mathrm{ext}$ (Fig.~\ref{fig:phase}).  In the absence of any interlayer coupling ($J_C$=0) and external magnetic field ($B_\mathrm{ext}$=0), the ground state of the system is made of spin spirals with $q$ vector $|\bm{q}_\mathrm{B,T}|$=$D/J$. For our choice of parameters, the optimal wavelength of such a spin spiral is $L_D$=$2\pi/|\bm{q}_L|$=31.5 which is also reflected in our numerical simulation. When viewed together as a bilayer, the superimposed spirals create a checkerboard pattern and therefore we call this phase \textit{checkerboard} phase (Fig.~\ref{fig:phase}d-f) (Also, see Supplementary Note I). This phase remains the ground state for small $J_C$ ($J_\mathrm{C}/J \lesssim 0.01$).  It is important to note that here we have the \textit{coexistence} of $q$-states (layer-by-layer), which should not be confused with the multi-$q$ \textit{approximation} of periodic magnetic textures~\cite{Okubo2012}. The feature that distinguishes this checkerboard phase from the earlier monolayer checkerboard configuration formed by multi-$q$ spin waves \cite{Hayami2023}, is the presence of points where the top and bottom layers are aligned against the direction favoured by the interlayer exchange (marked by orange circles in Fig.~\ref{fig:phase}b,d-f). We call these points \textit{anti-aligned} points. In the subsection \textit{Linked skyrmion}, we will show that these points  can be characterised as topological defects which play a crucial role in the formation of \textit{linked skyrmions}, a  configuration where multiple skyrmions are \textit{linked} together by such topological point defect. With the increase of interlayer coupling, such points become less favourable and the ground state is made of spin spirals where the $(\bm{q}_\mathrm{T} \rightarrow \bm{q}_\mathrm{B}) || \bm{e}_x \pm \bm{e}_y$. With the increase of the external magnetic field, the ground state is gradually made of skyrmion lattice. Note that, due to the anisotropic DMI, skyrmions in each layer have a small elongation (marked by white arrows in Fig.~\ref{fig:phase}g). As a result the skyrmions and therefore the skyrmion lattice itself possess a two-fold rotation symmetry, rather than conventional six fold rotation symmetry observed in a hexagonal skyrmion lattice \cite{Okubo2012, Science_2009}. Unlike the checkerboard or the spiral phase, the skyrmion lattice phase can be characterised by a non-zero topological charge (Eq.~\ref{Q}) where the skyrmion domain is the unit cell of the lattice (denoted by the dashed line in Fig.~\ref{fig:phase}g). For convenience, we define a layer-resolved topological charge with representative formulas for the continuum limit
\begin{eqnarray}
Q_{\mathrm{T},\mathrm{B}} = \frac{1}{4\pi} \int  dx \ dy \  \mathbf{m}^{\mathrm{T},\mathrm{B}} \cdot[\partial_{x} \mathbf{m}^{\mathrm{T},\mathrm{B}} \times\partial_{y} \mathbf{m}^{\mathrm{T},\mathrm{B}}],
\label{QL}
\end{eqnarray} 
while for our discrete problem, we calculate the $Q_{\mathrm{T},\mathrm{B}}$ values accordingly using the Berg-L\"{u}scher simplicial method~\cite{Berg1981}. In case of the skyrmion lattice, the integration is done over a finite region marked by dashed line in Ref.\,\ref{fig:phase}g which gives $Q_{\mathrm{T}}$=$Q_{\mathrm{B}}$=-1. 

Along with the skyrmion lattice, a large variety of complex metastable states with arbitrarily large topological charges can also coexist within the same parameter region. These states are broadly classified into two classes - (i) Linked skyrmions and  (ii) Skyrmion bags and k$\pi$-skyrmions. In case of a linked skyrmion one can have both $Q_{\mathrm{T}} \neq Q_{\mathrm{B}}$ and $Q_{\mathrm{T}}$=$Q_{\mathrm{B}}$ with one or multiple point defect whereas for the k$\pi$-skyrmion and skyrmion bags always have $Q_{\mathrm{T}}$=$Q_{\mathrm{B}}$ and no point defect. These are elaborated in the subsections \textit{Linked skyrmion} and \textit{Skyrmion bags and $k\pi$-skyrmions}

\subsection{Linked skyrmion}

The transition from the checkerboard phase to the skyrmion lattice phase is achieved by increasing the interlayer coupling and applying an external field. This makes the \textit{anti-aligned} points energetically unfavourable. However, survival of an \textit{anti-aligned} point leads to the formation of a magnetic texture where one can have $Q^\mathrm{T} \neq Q^\mathrm{B}$ which we call a \textit{linked skyrmion} (Fig.~\ref{fig:link}). Such configurations are observed over a wide range in the phase space where skyrmion lattice is observed (See Supplementary Note I and II). In Fig.~\ref{fig:link} we use $J_C/J$=0.02 and $\mu B_{ext}/J$=0.007 which are the same values used to stabilise skyrmion lattice (Fig.~\ref{fig:phase}g). Such texture can have any number of skyrmions in each layer resulting in an arbitrarily large topological charge.

\textbf{Topological nature of the skyrmion links: } At the centre of each skyrmion link, lies an \textit{anti-aligned} point. As a result, the average magnetisation of the bilayer in the surrounding region tends to be zero. Thus, the texture in Fig.~\ref{fig:link}a can be considered to contain a point defect. Similar to the Bloch point~\cite{Malozemoff1979} - a magnetic point defect in bulk crystals - the \textit{anti-aligned} points can also be characterized by a topological charge, $Q^\mathrm{D}$ which differs from the skyrmion topological charge (see Supplementary Note III). Here, the topological charge of such a point defect can be expressed in terms of the skyrmion topological charges in the top and bottom layers as
\begin{equation}
    Q^\mathrm{D} = Q^\mathrm{T} - Q^\mathrm{B}.
    \label{QPmain}
\end{equation}
For the linked-skyrmions, therefore, it is specifically the difference of $Q^\mathrm{T}$ and  $Q^\mathrm{B}$  that serves as a meaningful classifying quantity.

\subsection{Skyrmion bags and $k\pi$-skyrmions}

\begin{figure}[h!]
\centering
\includegraphics[width=\linewidth]{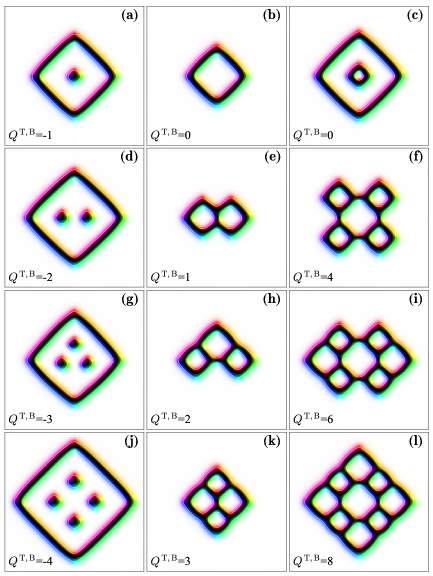}
\caption{Different composite skyrmions without a point defect. (a-c) $k\pi$-skyrmions and (d-l) skyrmion bags with different topological charges stabilised at ($J_C$=0.02$J$, $\mu B_\mathrm{ext}$=0.007$J$) in a 400$\times$400 mesh. The figures present the total magnetisation of both layers. The legends show the topological charge from the top and bottom layers.}
\label{fig:bags}
\end{figure}

The other class of configuration present in the same region of the phase space contains textures with $Q^\mathrm{T}$=$Q^\mathrm{B}$ and consequently $Q^\mathrm{D}$=0 (Fig.~\ref{fig:bags}). We use $J_C/J$=0.02 and $\mu B_{ext}/J$=0.007 for comparing them with skyrmion lattice and linked skyrmions. These textures can be further divided into two subclasses namely skyrmion bags~\cite{Rybakov2019, Foster2019} and so-called $k\pi$-skyrmions~\cite{Bogdanov1999}. Note that, unlike the linked skyrmion, here the total topological charge can be both positive or negative as well as zero.

\subsection{Material realisation}

\begin{figure}[ht!]
\includegraphics[width=0.48\textwidth]{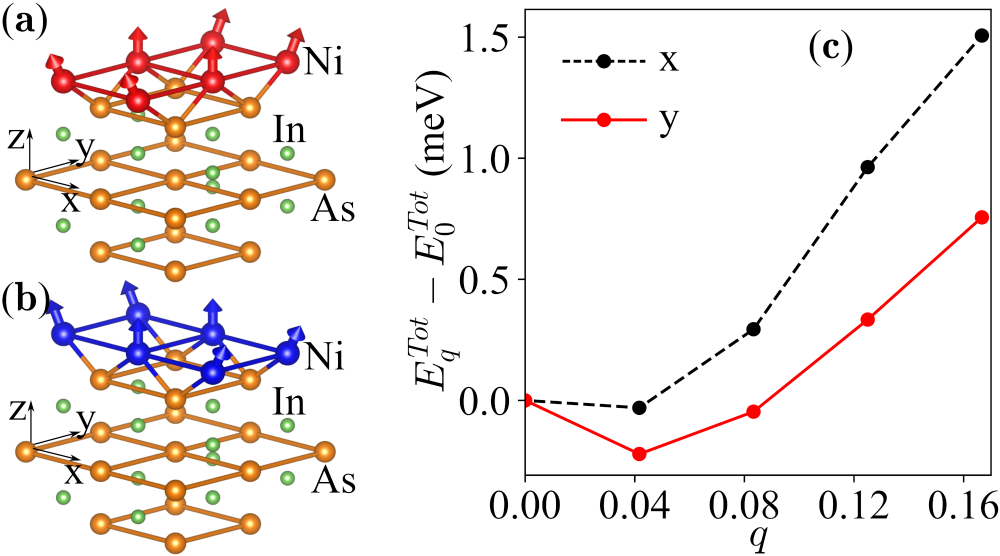}
\caption{Material realisation of the bilayer model. The thin-film structure of the Ni/InAs(001) interface for (a) top and (b) bottom layer. Red and blue spheres denote Ni (magnetic) for the top and bottom layers, and yellow spheres denote In (the main source of SOC). As is shown in green. (c) Total energy of spin spiral for the top layer (a), where the red and black dashed lines represent the spiral along $y$ and $x$ direction on the top layer.}
\label{fig:NiInAs}
\end{figure}

We have conducted a thorough first-principles study to identify a suitable material for this system. Our studies show that the thin-film structure of Ni/InAs(001) can be an ideal candidate which possesses all the relevant symmetry and necessary DMI (Fig.~\ref{fig:NiInAs}). The exchange and DMI along $x$ and $y$ direction are obtained from the total energy of the spin spiral with different $q$ vectors (Fig.~\ref{fig:NiInAs}b). Although our calculation shows the existence of finite exchange and DMI beyond the first nearest neighbours (see Supplementary Note IV for details), the dominant contribution comes from the first nearest neighbours. Considering the leading order contribution, for the top layer we find that $J_x$=1.4\,meV, $D_x$=0\,meV and $J_y$=1.6\,meV, $D_y$=0.1\,meV where the suffix denotes the direction of the bond. Due to the symmetry, the bottom layer would have the same magnitude of the $J$ and $D$ parameters with $x$ and $y$ being interchanged. The atomic magnetic moment of Ni is found to be 0.55\,$\mu_B$. Our numerical simulations show that with these magnitude of exchange and DMI one can stabilise a linked skyrmion similar to that shown in Fig.~\ref{fig:link}b with a magnetic field of 70\,mT over a wide range of $J_C$ ($0.002 < J_C/J < 0.005$) with a typical dimension of 130\,nm$\times$45\,nm.

\section{Conclusion}
In this work, we introduce a shifted magnetic bilayer that hosts a wide range of magnetic configurations which can be obtained by tuning the interlayer exchange coupling and an external magnetic field. A characteristic feature of the ground state at weak interlayer coupling and external field is the presence of special regions in which the interlayer magnetic alignment is opposite to the direction favoured by the interlayer exchange. We call these points \textit{anti-aligned} points, which play a crucial role in connecting multiple solitons creating a class of complex magnetic textures with arbitrarily large topological indices, referred to here as \textit{linked skyrmions}. A homotopy group analysis demonstrates that these points are topologically stable. In addition to linked skyrmions, our model also supports conventional skyrmion bags and $k\pi$-skyrmions, both of which can also host arbitrarily large topological charges. However, linked skyrmions provide a higher topological charge density, which can significantly enhance their transport features, such as a larger effective Hall angle, as well as increased  non-linearity of their dynamical response. These features make them more promising candidates for skyrmion based computing compared with conventional skyrmions. Finally, based on first-principles calculations, we propose suitable materials in which such solitons can be observed. As a concrete example, we identify a Ni/InAs(001) thin-film system that possesses all required symmetries and interactions. The interlayer coupling can be tuned by controlling the thickness of the InAs layers and also by inserting a suitable non-magnetic spacer layer. Owing to the generality of our theoretical framework, a broad class of materials is expected to satisfy the necessary symmetry conditions to realise these phenomena. Our results therefore provide a route towards engineering higher-order topological solitons in realistic magnetic systems. 

\section{Method}

\subsection{Atomistic simulation}
The atomistic simulation is done with the atomistic simulation code Spirit \cite{SPIRIT}. The optimised solutions are obtained by minimising the energy with a threshold of 10$^{-10}J$. In case of periodic configurations shown in Fig.~\ref{fig:phase}d-i the optimum system size is obtained by varying the dimension and finding the dimension corresponding to the minimum energy for each set of parameters. For the isolated solitons (Figs.~\ref{fig:link}, \ref{fig:bags}) we use $D$=0.2$J$, $J_C$=0.02$J$ and $\mu B_\mathrm{ext}$=0.007$J$ and a mesh size of 400$\times$400 to avoid finite size effects.

\subsection{Homotopy-group analysis}

Following the approach proposed in Ref.~\onlinecite{RybakovEriksson2022, Rybakov2025} (See Supplementary Note III for details), we first define an effective space for the order parameter by analysing the most unfavourable spin orientations as
\begin{eqnarray}
X = \left\lbrace (\bm{m}^\mathrm{T},\bm{m}^\mathrm{B}) ;\, \bm{m}^\mathrm{T}  \in \mathbb{S}^2,\, \bm{m}^\mathrm{B}  \in  \mathbb{S}^2,\, \bm{m}^\mathrm{T} \neq -\bm{m}^\mathrm{B} \right\rbrace.
\label{space1}
\end{eqnarray}

Note that $X$ is the homotopy equivalent to $\mathbb{S}^2$. Considering the first homotopy group $\pi_1(X)$=$\pi_1(\mathbb{S}^2)$=0, one can see that there are no stable vortices in this system. Each layer can have ordinary skyrmions corresponding to the map $I^2 / \partial I^2 \rightarrow \mathbb{S}^2$, where $I^2$ is a rectangle and $\partial I^2$ is its perimeter. The corresponding topological charges of the skyrmions in each layer are given by
\begin{equation}
Q^\mathrm{L} = \frac{1}{4\pi} \int dS\ \mathbf{F}^\mathrm{L}\cdot\hat{\mathrm{e}}_z, \label{QL1}
\end{equation}
where $F_z^\mathrm{L}$=$\bm{m}^\mathrm{L} \cdot (\partial_x \bm{m}^\mathrm{L} \times \partial_y \bm{m}^\mathrm{L})$ and L=T,B represents the top and the bottom layers. In addition to that, the system can have stable point defects corresponding to the second homotopy group $\mathbb{S}^2 \rightarrow \mathbb{S}^2,\, \pi_2(\mathbb{S}^2) = \mathbb{Z}$. The preimage of the map is a closed surface of containing a point defect (See Fig.~\ref{fig:link}). The topological charge of this point defect can be defined by the Kronecker integral
\begin{eqnarray}
Q^\mathrm{D} = \frac{1}{4\pi} \oiint dS\ \bm{F} \cdot \hat{\bm{n}}, \label{QD1}
\end{eqnarray} 
where $\hat{\bm{n}}$ is the unit normal to the surface and $\bm{F}$ is the vector curvature. For a trapezoidal volume around the point defect, we have $\hat{\bm{n}}^\mathrm{T}=-\hat{\bm{n}}^\mathrm{B}$=$\hat{e}_z$. Using Eq.~\ref{QL1} and keeping in mind that the net contribution from the side surfaces vanishes, we obtain
\begin{eqnarray}
Q^D = Q^\mathrm{T} - Q^\mathrm{B}\, ,
\end{eqnarray}   
which represents the topological point defect as the difference of the stable skyrmion topological charges of the individual layers.

\subsection{First-principles calculations}
The first-principles calculations were performed using the \textit{ab initio} code \textsc{FLEUR}\cite{Fleur}. Here we use a symmetric thin film structure (see Supplementary Note IV for details). By placing a single layer of Ni in a square lattice on top of the (001) plane of InAs, such that the square lattice is shifted by half the lattice vector along the $y$ (for top layer, Fig.~\ref{fig:NiInAs}a) or the $x$ (for bottom layer, Fig.~\ref{fig:NiInAs}b) axis with respect to the square formed by the adjacent As atoms, the resulting crystal symmetry will produce the same orientation of the D-vector as shown in Fig.~\ref{fig:bilayer}. 
For the structure in Fig.~\ref{fig:NiInAs}, the most stable in-plane lattice constant near the InAs lattice constant was determined (4.268\,\AA) by relaxation of internal coordinates along the $z$-axis such that the force on each atom is below 10$^{-6}$\,Hartree/Bohr. The result shows that the structure remains stable with Ni bound to InAs without detachment. After the structural relaxation, we calculate the energy of the spin spiral using the generalised Bloch theorem \cite{Heide2009}. For our calculation we use a 48$\times$48$\times$1 k-mesh. The total energy for different values of $q$ along $x$ and $y$ axis is shown in Fig.~\ref{fig:NiInAs}b. To obtain $J$ we calculate the energy of the spin spiral along direction $\hat{i}$ without SOC and fit it with $E(q_i)$=$E(-q_i)$=$-$$E_0$ $-$$\sum_n J_i^n \cos(2n \pi q_i)$ where $J_i^n$ is the Heisenberg exchange between $n^{th}$ nearest neighbour along $\hat{i}$. For the DMI we calculate the energy due to SOC and fit it with $E_{\mathrm{SOC}}(q_i)=-E_{\mathrm{SOC}}(-q_i)=-\sum_n D_i^n \sin(2n \pi q_i)$ where $D_i^n$ is the DMI vector between $n^{th}$ nearest neighbour along $\hat{i}$ direction (see Supplementary Note IV for details).

\section{Data availability}
The datasets generated during and/or analysed during the current study are available from the corresponding author on reasonable request.

\section{Acknowledgement}
SG is co-funded by the European Union (Physics for Future – Grant Agreement No. 101081515). This work is supported by the Ministry of Education, Youth and Sports of the Czech Republic through the e-INFRA CZ (ID:90254). This project has received funding from the European Research Council under the European Union's Horizon 2020 Research and Innovation Programme (Grant No.~856538 - project ``3D MAGiC''). HK acknowledges computing time on the supercomputer JURECA \cite{Thornig2021-mu} at Forschungszentrum Jülich under grant no. JIFF38.
FNR and OE acknowledge support from the Swedish Research Council (Grant No.~2023-04899).

\section{Author contributions}
SG, HK and SB have initiated the project. SG has established the model and  obtained different solutions. HK has conducted the symmetry analysis,  identified suitable material candidate and carried out the first-principles calculation with GB. FR has carried out the homotopy group analysis. MS has helped with the atomistic simulation. VK has helped with the phase diagram. SB, OE and NSK have contributed in analysing the results. SG has written the manuscript with feedback from others. All co-authors have discussed the results and contributed equally.

\section{Competing interests}
The authors declare no competing interests.

\bibliography{main}

\pagebreak

\begin{center}
\textbf{\large Supplementary Materials}
\end{center}

\section*{Supplementary Note I: \\ Initial states for different configurations}

\begin{figure}[h!]
\centering
\includegraphics[width=0.45\textwidth]{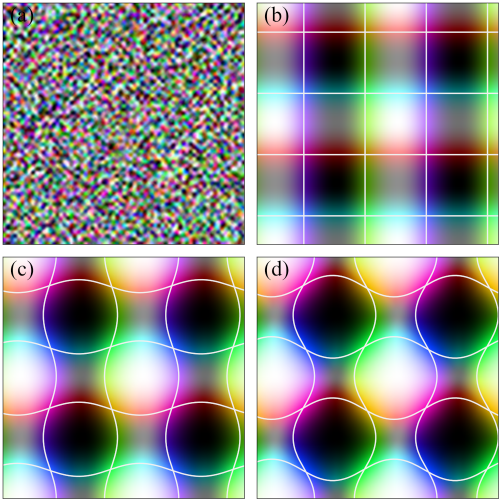}
\caption{Different checkerboard phases obtained from the random distribution. Here we use $D/J$=0.2 and $B_{ext}$=0 and a 70$\times$70 mesh. Figures show the bilayer configurations as shown in Fig.~2d-f in the main text. (a) Initial complete random distribution. Checkerboard configuration for (a)$J_C/J$=0, (c)$J_C/J$=0.006 and (d)$J_C/J$=0.012. (a), (b) and (c) are used as initial states to obtain (b), (c) and (d).}
\label{fig:initCH}
\end{figure}

In this section we provide details of the initial states used to obtain different configurations. For the checkerboard phase one can start from a complete random state Suppl.\ Fig.~\ref{fig:initCH}a) and obtain two mutually perpendicular spin spirals in absence of any interlayer coupling (Suppl.\ Fig.~\ref{fig:initCH}b). Using this state as initial state one can obtain the checkerboard phase at higher interlayer coupling (Suppl.\ Fig.~\ref{fig:initCH}c). This step can be repeated successively to obtain the checkerboard phase at higher interlayer coupling (Suppl.\ Fig.~\ref{fig:initCH}d). For the spin spiral phase (Fig.~2h,i in the main text) one can star from an initial spin spiral state where the $q$ vectors will be parallel in both layers.

\begin{figure}[h!]
\centering
\includegraphics[width=0.48\textwidth]{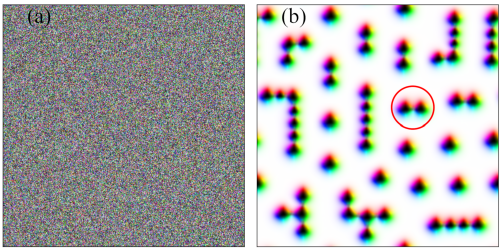}
\caption{Different linked skyrmions obtained at $D/J$=0.2, $\mu B_{ext}/J$=0.01 and $J_C/J$=0.02 on a 400$\times$400 mesh. Here we plot the average magnetisation of the two layers. (a) shows the initial random state and (b) shows the final configuration. The red circle denotes a specific linked skyrmion that has been isolated and used to study its stability.}
\label{fig:R2}
\end{figure}

One can also start from a completely random configuration to obtain different linked-skyrmions. In this case, multiple linked skyrmions with different topological charge can appear simultaneously (Suppl.\ Fig.~\ref{fig:R2}). Here, we use a comparatively large magnetic field to obtain a higher density of the linked skyrmions. One can isolate a specific linked skyrmion from this collection to use it for further analysis.

\begin{figure}[h!]
\centering
\includegraphics[width=0.48\textwidth]{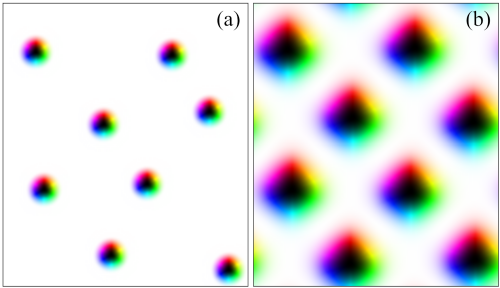}
\caption{Skyrmion lattice from the random distribution of skyrmions. Here we use $D/J$=0.2, $\mu B_{ext}/J$=0.007 and $J_C/J$=0.02 on a 130$\times$150 mesh. (a) Initial configuration and (b) skyrmion lattice ground state obtained from (a).}
\label{fig:sklat}
\end{figure}

If one starts from an initial state that consists of multiple isolated skyrmions, one can obtain the skyrmion lattice as the ground state. In this case, we choose a rectangular mesh to accommodate a near-triangular lattice of skyrmions (Suppl.\ Fig.~\ref{fig:sklat}).

\begin{figure}[h!]
\centering
\includegraphics[width=0.48\textwidth]{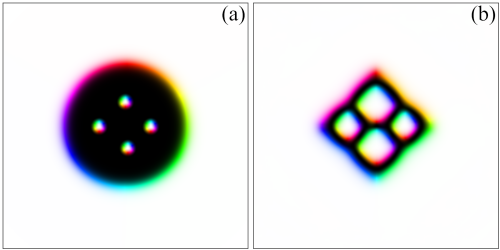}
\caption{Formation of skyrmion bag at $D/J$=0.2, $\mu B_{ext}/J$=0.01 and $J_C/J$=0.02 on a 400$\times$400 mesh. (a) shows the initial configuration and (b) shows the final configuration.}
\label{fig:bag}
\end{figure}

One can similarly obtain the skyrmion bags and $k\pi$ skyrmions by combining multiple skyrmions and anti-skyrmions as an initial state (Suppl.\ Fig.~\ref{fig:bag}).

\section*{Supplementary Note II: \\Stability of the linked skyrmion}

\begin{figure}[h!]
\centering
\includegraphics[width=0.45\textwidth]{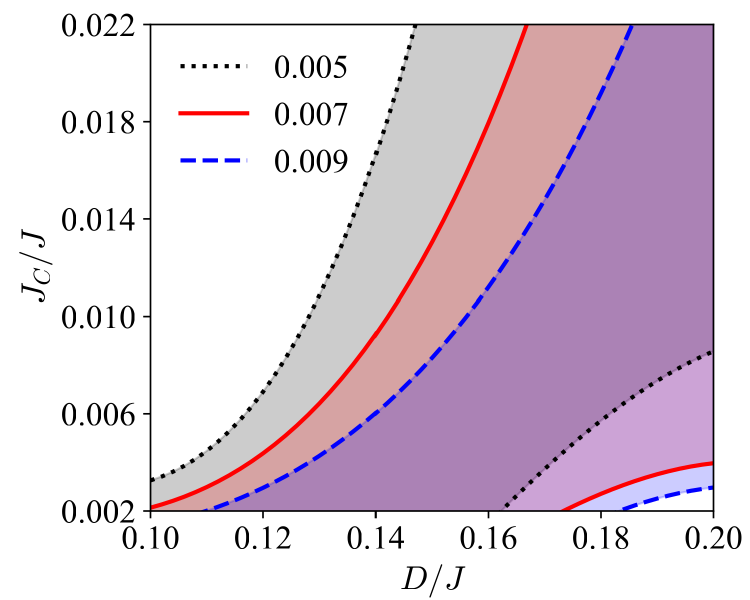}
\caption{Region of stability for the linked skyrmions. (a)Black dotted, (b)red and (c) blue dashed lines show the boundary of the region of stability for $\mu B_{ext}/J$=0.005 (black dotted), 0.007 (red, solid) and 0.009 (blue, dashed) respectively. The regions are denoted by the transparent colours.}
\label{fig:jd}
\end{figure}

Finally, we demonstrate stability of the linked skyrmions, which depends on all material parameters such as the DMI ($D$), interlayer coupling $J_C$ and external magnetic field ($B_{ext}$). Note that both $J_C$ and $B_{ext}$ favour ferromagnetic alignment. Therefore for stronger $J_C$ and $B_{ext}$, the linked skyrmions require stronger $D$ to stabilise (Suppl.\ Fig.~\ref{fig:jd}).

\begin{figure}[ht!]
\centering
\includegraphics[width=0.48\textwidth]{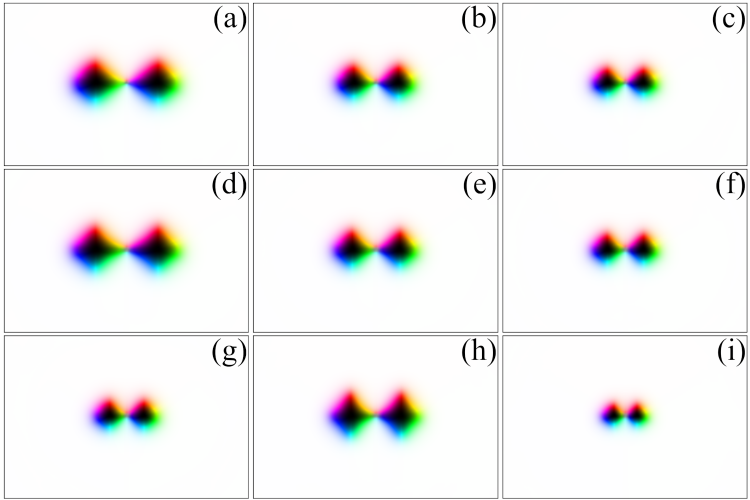}
\caption{Stability of linked skyrmion at different parameters. Here we use a 400$\times$400 mesh for calculation and plot the converged configuration on a 300$\times$200 region. The parameters used for different configurations are (a)$J_C/J$=0.03, $\mu B_{ext}/J$=0.006, $D/J$=0.20, (b)$J_C/J$=0.03, $\mu B_{ext}/J$=0.007, $D/J$=0.20, (c)$J_C/J$=0.03, $\mu B_{ext}/J$=0.008, $D/J$=0.20, (d)$J_C/J$=0.02, $\mu B_{ext}/J$=0.006, $D/J$=0.20, (e)$J_C/J$=0.02, $\mu B_{ext}/J$=0.007, $D/J$=0.20, (f)$J_C/J$=0.02, $\mu B_{ext}/J$=0.008, $D/J$=0.20, (g)$J_C/J$=0.02, $\mu B_{ext}/J$=0.007, $D/J$=0.18, (h)$J_C/J$=0.02, $\mu B_{ext}/J$=0.007, $D/J$=0.22 and (i)$J_C/J$=0.02, $\mu B_{ext}/J$=0.01, $D/J$=0.2. (i) is extracted from Suppl.\ Fig.~\ref{fig:R2}b (marked with red circle) and used as initial state to obtain the converged configurations (a)-(h).}
\label{fig:C2}
\end{figure}

To demonstrate the stability of the linked skyrmion at different parameter values, we choose a linked skyrmion obtained at $D/J$=0.2, $\mu B_{ext}/J$=0.01 and $J_C/J$=0.02 (marked with the red circle in Suppl.\ Fig.~\ref{fig:R2}b) as the initial configuration and vary all these three parameters. Our results (Suppl.\ Fig.~\ref{fig:C2}) show that the linked skyrmions are stable over a large domain of parameters.

\section*{Supplementary Note III: \\Homotopy-group analysis \label{app:Homotopy}}
\subsection{The fundamental group}
First, let us identify the fundamental group of possible topological defects in our system.
For every layer considered separately, the order parameter space is $\mathbb{S}^2$, and no stable vortices are expected since  $\pi_1(\mathbb{S}^2)=0$. 
However, the situation is less evident in the case of two monolayers coupled through the interlayer exchange.
Following the approach proposed in Ref.~\cite{RybakovEriksson2022}, we define an effective space for the order parameter constrained by the model Hamiltonian.
For spins from the top and bottom layers and interacting ferromagnetically, the most energetically unfavorable state is antiparallel. 
Hence, the effective order parameter space along the loop surrounding these defects is 
\begin{equation}
X\!=\!\{ (\bm{m}^\mathrm{T},\bm{m}^\mathrm{B}) ;\, \bm{m}^\mathrm{T}\!\! \in \!\mathbb{S}^2,\, \bm{m}^\mathrm{B}\!\! \in \!\mathbb{S}^2,\, \bm{m}^\mathrm{T}\!\neq-\bm{m}^\mathrm{B} \}.
\label{space}
\end{equation}

The space $X$ is homotopy equivalent to the $\mathbb{S}^2$, which can be explicitly illustrated by the deformation retract~\cite{Kosniowski1980}.
In our case, the space $X$ allows retraction to $\mathbb{S}^2$ and this map can be explicitly written as follows:
\begin{equation}
(\bm{m}^\mathrm{T},\bm{m}^\mathrm{B}) \mapsto 
\left(  
\frac{\bm{m}^\mathrm{T} + \tau\bm{m}^\mathrm{B}}{|\bm{m}^\mathrm{T} + \tau\bm{m}^\mathrm{B}|},  
\frac{\bm{m}^\mathrm{B} + \tau\bm{m}^\mathrm{T}}{|\bm{m}^\mathrm{B} + \tau\bm{m}^\mathrm{T}|}
\right), 
\end{equation}
where $\tau\in[0,1]$, while the denominators are never zeroing due to the last inequality in~(\ref{space}).
As it follows from the above, the fundamental (the first) homotopy group of $X$ is trivial:
\begin{equation}
\pi_1(X) \cong \pi_1(\mathbb{S}^2) = 0. 
\end{equation}
This implies that the model does not support topologically stable vortices, which entirely agrees with our observation in the numerical experiment.

\begin{figure}[h!]
\centering
\includegraphics[width=0.5\textwidth]{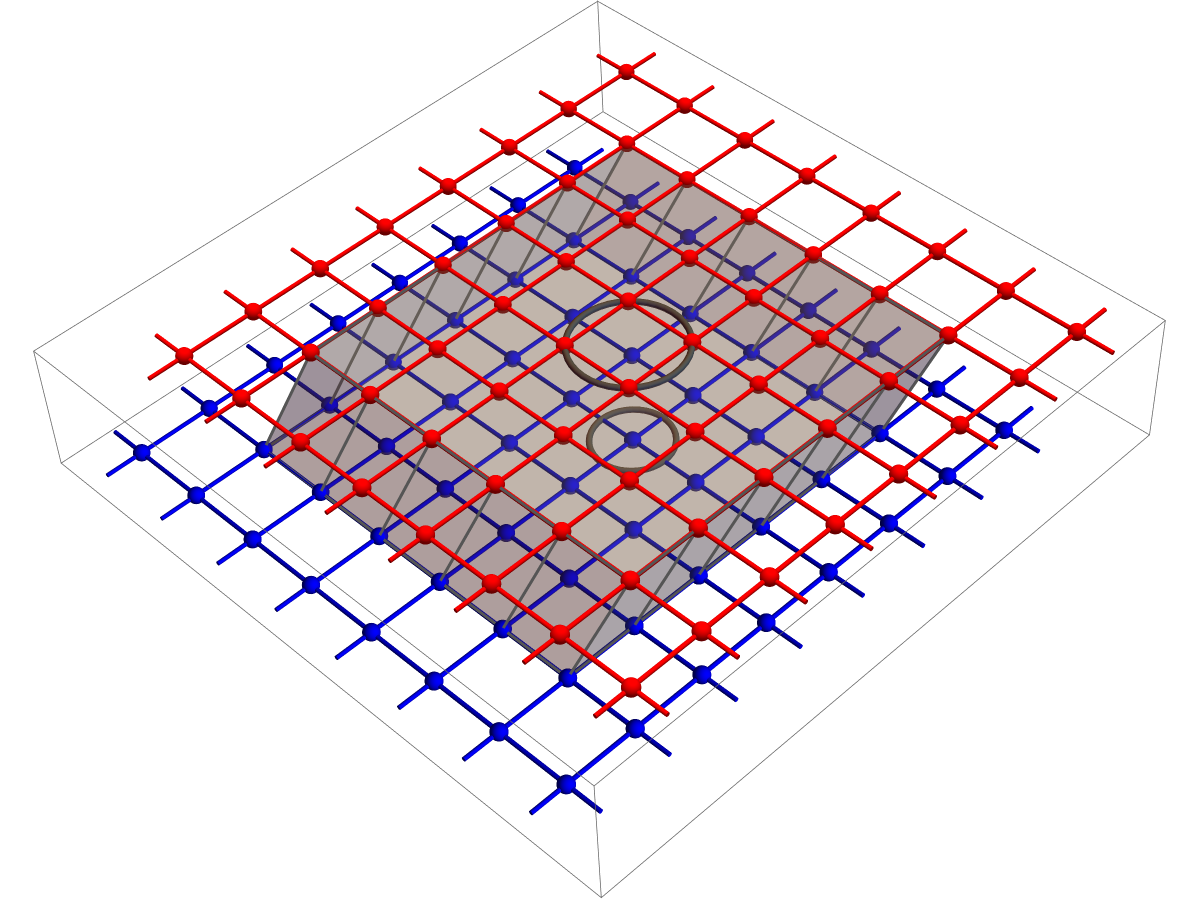}
\caption{Sketch of the volume of integration for Eq.~\ref{QP} that encloses a point defect (marked by circle). At the edges, the top and bottom layer are parallel (See Fig3a, bottom pannel in the main text).
\label{fig:prism}
}
\end{figure}

\subsection{Higher order homotopies}
Assuming that the Heisenberg exchange represents, as usual for most of the ferromagnets, the leading energy term, one should expect near-continuous variations of spins from site to site almost everywhere, except for point defects.
Accordingly, the classifying map is 
\begin{equation}
f:\ \mathbb{S}^2 \rightarrow \mathbb{S}^2. 
\label{map_f}
\end{equation}
The preimage of the map~$f$ in~(\ref{map_f}) represents a closed surface containing a point defect and, in accordance with the geometry of the lattice, allowing the shape of a prism [Suppl.\ Fig.~\ref{fig:prism}].
%
%
If we denote by $\hat{\mathbf{n}}$ the outer normal for any face of the prism, then the corresponding topological charge of a point defect can be determined by Kronecker integral: 
\begin{equation}
Q_\text{D} = \frac{1}{4\pi} \oiint dS\ \mathbf{F} \cdot \hat{\mathbf{n}},
\label{QP}
\end{equation}
where 
\begin{equation}
\mathbf{F} = 
\left( 
\begin{array}{c}
\mathbf{m}\cdot[\partial_{y}\mathbf{m}\times\partial_{z}\mathbf{m}]\\ 
\mathbf{m}\cdot[\partial_{z}\mathbf{m}\times\partial_{x}\mathbf{m}]\\ 
\mathbf{m}\cdot[\partial_{x}\mathbf{m}\times\partial_{y}\mathbf{m}]   
\end{array}
\right)
\end{equation}
\noindent is the vector of curvature~\cite{Aminov_1969, Zheng2023}
and $x,y,z$ are local right-handed Cartesian coordinates, defined in each face of the prism such that $\hat{\mathbf{r}}_3\cdot\hat{\mathbf{n}}=1$. 
Equivalently, $Q_\mathrm{D}$ can be calculated by the Berg and L\"{u}scher method~\cite{Berg1981} on the triangulated prism surface.

The topological invariant~\eqref{QP} is similar to that of a Bloch point~\cite{Malozemoff1979}, a magnetic point defect in the bulk systems. 
The key difference is that in bulk, the convex surface $S$, over which the integration is performed, can formally be extended to a sphere with the diameter tending to infinity, while, in our bilayer system, $S$ is always confined by two layers.

Remarkably, the topological index of such a point defect, besides Eq.~\eqref{QP}, can also be expressed in terms of skyrmion topological charges of the top layer, $Q^\mathrm{T}$, and the bottom layer, $Q^\mathrm{B}$ given by
\begin{eqnarray}
Q_{\mathrm{T},\mathrm{B}} = \frac{1}{4\pi} \int  dx \ dy \  \mathbf{m}^{\mathrm{T},\mathrm{B}} \cdot[\partial_{x} \mathbf{m}^{\mathrm{T},\mathrm{B}} \times\partial_{y} \mathbf{m}^{\mathrm{T},\mathrm{B}}],
\label{QTB}
\end{eqnarray} 
The skyrmion classifying map corresponds to the second homotopy group of the sphere:
\begin{equation}
g:\ I^2 / \partial I^2 \rightarrow \mathbb{S}^2,
\label{map_g}
\end{equation}
where $I^2$ is a rectangle in one of the layer and $\partial I^2$ -- its perimeter. Note that the domain can be chosen in any other way while remaining homeomorphic to a rectangle. 
The one-point compactification of the rectangle is possible due to the fixed magnetization value along its perimeter, $g(\partial I^2)=\mathbf{m}_0$. 
The preimage of the map~$g$ is homeomorphic to the two-sphere, $I^2 / \partial I^2 \cong \mathbb{S}^2$, so the Eq.~\eqref{map_g} is effectively a map from one two-sphere to another as in~\eqref{map_f}.
The corresponding topological charges of skyrmions can be found again using the Kronecker integral (or Berg and L\"{u}scher method) but on a flat domain: 
\begin{equation}
Q_{L} = \frac{1}{4\pi} \int dS\ \mathbf{F}\cdot\hat{\mathrm{e}}_z, \label{QL2}
\end{equation}
where $L$ defines the top or bottom layer (T or B).
Note that formulas \eqref{QL2} and \eqref{QTB} are equivalent.

From ~(\ref{QP}) and~(\ref{QL2}) follows the connection between the topological charge of the point defect and skyrmion charges in the top and bottom layers:
\begin{equation}
Q_\mathrm{D} = Q_\mathrm{T} - Q_\mathrm{B}.
\label{QP1}
\end{equation}
The contribution from the side facets of a prism vanishes in Eq.~(\ref{QP1}) because of fixed magnetization,~$\mathbf{m}_0$, at the periphery  (see Fig.~3a bottom panel in the main text).
In other words, the magnetization at the side facets of a prism represents a collinear state with $\mathbf{F}=0$.

\section*{Supplementary Note IV: \\First-principles study \label{app:DFT}}

\begin{figure}[h!]
\centering
\includegraphics[width=0.45\textwidth]{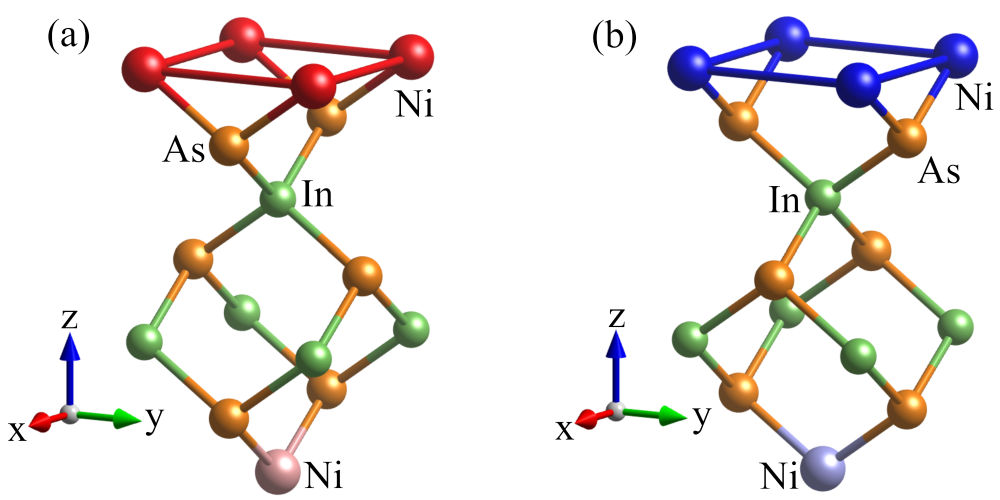}
\caption{Lattice structure of symmetrised NiInAs thin film. Unit cell of the (a) top and (b) bottom layer where the (a)light red and (b)light blue atoms represent a Ni atom with magnetisation perpendicular to all the (a)red and (b)blue atoms respectively.}
\label{fig:NiInAs1}
\end{figure}

Here we consider the interface between a layered structure of fcc InAs as the source of SOC and Ni (Suppl.\ Fig.~\ref{fig:NiInAs1}). For simplicity, we consider three layers of In and single layers of Ni. We add an additional layer of Ni (light red and light blue spheres in Suppl.\ Fig.~\ref{fig:NiInAs1}a,b) at the bottom to symmetrise the structure. The magnetisation on the bottom layer is kept perpendicular to the top layer and therefore does not contribute to the interlayer exchange energy. The relaxed structure possesses an in-plane lattice vector 4.268\,\AA.

The effective magnitudes of exchange ($J$) and DMI ($D$) are calculated from the dispersion of the spin spiral for Ni on the top surface. The total energies of the spin spirals with different $q$ vectors are calculated using the generalised Bloch theorem with $48\times48\times1$ k-points. For $\bm{q}=(0,0,0)$, the magnetic moment of Ni on the top surface is oriented in the $x$ direction in the plane, while the magnetic moment of Ni on the bottom surface is oriented in the $y$ direction. Thus, only the $q$-dependence of the magnetic moment of Ni on the top surface can be considered since the magnetic moments of the upper and lower Ni are perpendicular to each other. To obtain the exchange parameter ($J$), we calculate the dispersion for the spin spiral without SOC which results in a symmetric dispersion ($E(q)$=$E(-q)$). The exchange coupling can be obtained by exploiting the relation $E(q)=-\sum_n J_n \cos(2\pi n q)+E_0$, where the suffix $n$ corresponds to the coupling between the $n^{th}$ nearest neighbour. To obtain the DMI, we introduce SOC and calculate the energy due to SOC. This contribution is odd in $q$ ($E(q)$=-$E(-q)$) and different orders of DMI can be obtained by fitting it with $E(q)=-\sum_n D_n \sin(2\pi n q)$. The values obtained for $J$ and $D$ along $x$ and $y$ direction considering upto 8 nearest neighbours are shown in Suppl.\,Table\,\ref{table:JD}.

\begin{table}[h!]
\centering
\begin{tabular}{|c|r|r|r|r|r|r|r|r|}
\hline
      & 1~~  & 2~~  & 3~~  & 4~~  & 5~~  & 6~~  & 7~~  &  8~~ \\ \hline
$J_x$ & 1.41 & 0.38 & 0.17 & 0.04 &-0.08 &-0.06 & 0.08 &-0.10 \\
$J_y$ & 1.61 & 0.37 & 0.09 &-0.12 &-0.23 & 0.03 & 0.36 &-0.35 \\
$D_x$ & 0.01 & 0.01 & 0.01 & 0.00 & 0.00 & 0.00 & 0.00 & 0.00 \\
$D_y$ & 0.09 & 0.06 & 0.05 & 0.00 &-0.01 &-0.00 & 0.01 &-0.00 \\ \hline
\end{tabular}
\caption{$J$ and $D$ parameter (in meV) along $x$ and $y$ direction.}
\label{table:JD}
\end{table}

\end{document}